\documentstyle[twocolumn,aps,prl,epsf,floats]{revtex}

\newcommand{\be}{\begin{eqnarray}}  
\newcommand{\ee}{\end{eqnarray}}

\newcommand{\jtlea}[1]{\label{#1}}
\newcommand{\jtleq}[1]{\label{#1}}

\newcommand{\jtcite}[1]{{\small{$^{\cite {#1}}$}}} 

\newcommand{\vect}[1]{\bf #1}

\newcommand{\cali}[1]{\cal #1}

\font\upright=cmu10 scaled\magstep1

\newcommand{\1}{\hbox{\upright\rlap{I}\kern -2.5pt 1}}
\def\square{\vcenter{\vbox{\hrule height.4pt
          \hbox{\vrule width.4pt height4pt
          \kern8pt\vrule width.4pt}\hrule height.4pt}}}

\begin{document}
\twocolumn[\hsize\textwidth\columnwidth\hsize\csname
@twocolumnfalse\endcsname

\typeout{--- Title page start ---}

\renewcommand{\thefootnote}{\fnsymbol{footnote}}


\title{The Jaynes-Gibbs principle of maximal entropy and
the non-equilibrium propagators of the $O(N)\;\phi^{4}$ theory at large $N$}
\author{P.Jizba, E.S.Tututi}
\address{DAMTP, University of Cambridge, 
Silver Street, Cambridge, CB3 9EW, UK}

\date{\today}

\maketitle

\begin{abstract}\noindent 
We   present  a   novel  procedure  for   calculating
non-equilibrium two-point Green's functions  in the  $O(N)\; \phi^{4}$
theory   at large $N$.   The non-equilibrium  density matrix $\rho$ is
constructed via the Jaynes-Gibbs  principle of maximal entropy  and it
is directly implemented  into  the  Dyson-Schwinger equations (DSE)  through
initial  value conditions.    In the large  $N$  limit   we perform an
explicit evaluation   of   two-point    Green's  functions   for  two
illustrative choices of $\rho$. 

\draft

\end{abstract}
\vskip1.5pc]



\renewcommand{\thefootnote}{\fnsymbol{footnote}}
\setcounter{footnote}{0}

\typeout{--- Main Text Start ---}

\section{Introduction}

In   the    present work we    show   a  particular   approach to  the
non-equilibrium  QFT dynamics based   in the Jaynes-Gibbs principle of
maximal entropy\jtcite{Jayn,Jayn2,Tol}.  In contrast to  other methods
in use\jtcite{makhlin,CHKMP,M1,M2,EJY}   the    Jaynes-Gibbs    method
constructs a   density   matrix  $\rho$ directly  from    the observed
macroscopic    quantities   (e.g.    pressure,  density    of  energy,
magnetisation,   particle  current,    local momentum,  local  angular
velocity,  ionisation rate (if   plasma is in   question), etc.).  The
$\rho$ is then implemented through the generalised Kubo-Martin-Schwinger
(KMS) conditions  into the dynamical  equations for Green's functions.
To keep complexity minimal we illustrate our  method on a paradigmatic
physical system  described  by $\lambda\phi^{4}$  theory with   $O(N)$
internal symmetry. The plan of this paper is as follows. 

In  Sec.II we review the  Jaynes-Gibbs  principle.  The  r\^ole of the
Shannon entropy\jtcite{Jayn,Sh} is emphasised.   Sec.III is devoted to
the construction of the DSE for QFT  systems away from equilibrium. We
use the  canonical formalism which appears  to be very natural in this
context.  In order to  reflect  the  density  matrix in the  dynamical
equations  we   show how   to   formulate  the  relevant  initial-time
conditions. The  DSE for two-point  Green functions are  worked out in
Sec.IV. With this mathematical setting, we take in Sec.V the large $N$
limit.  In the latter  case the DSE  for two-point Green functions are
decoupled   (virtually by chance). We explicitly   solve these for two
illustrative choices of the initial-time conditions.

\section{Initial conditions, Jaynes-Gibbs principle of maximal entropy}

In  this section we would  like to briefly  review   the
Jaynes-Gibbs principle  of   maximal  entropy (also  maximum   calibre
principle)\jtcite{Jayn,Jayn2,Tol,Gibbs}.    The   objective   of   the
principle   is to construct the  `most  probable' density matrix which
fulfils the constraints imposed by experimental/theoretical data.

The standard rules of statistical physics allows us to define
the expectation value via the density matrix $\rho$ as $
\langle \cdots \rangle = \mbox{Tr}(\rho \cdots)$,
with the trace running over a complete set of {\em physical}
states describing the ensemble in question at some initial time $t_{i}$.

The    usual  approaches\jtcite{CHKMP,M1,M2,EJY} trying   to determine
$\rho$  start with the     Liouville  equation and  hence  with    the
Schr{\"o}dinger picture.   Rather than following   this path, we shall
use the  Heisenberg picture instead. This will prove useful in
Sec.III. 

In order to determine $\rho$ explicitly we shall resort to the
Jaynes-Gibbs principle of maximal entropy\jtcite{Jayn,Jayn2,Tol,Gibbs}. 
The basic idea is `borrowed' from the information theory. Let us
assume that we have criterion of how to characterise the informative
content of $\rho$. The most ``probable'' $\rho$ is then selected out of
those $\rho$ which are consistent with `whatever' we know about the system
and which have the last informative content (Laplaces's principle of
insufficient reasons).

It remains to characterise the information content (measure) 
$I[\rho]$ of $\rho$. This was done by C.Shannon\jtcite{Sh} with the
result: $I[\rho] = \mbox{Tr}(\rho \; {\mbox{ln}}\rho)$.

The  density matrix\footnote{ It  can be shown\jtcite{BM}
that $-I[\rho]$ (also called the informative  entropy) equals (in base
2 of  logarithm) to the  expected number of binary (yes/no)
questions whose answer takes us from our  current state of knowledge to
the one certainty.} is then chosen to minimise $I[\rho]$. Note that no
assumption about the  nature of $\rho$ was made;  namely  there was no
assumption whether  $\rho$   describes equilibrium or  non-equilibrium
situation. To put more flesh on the bones,  let us rephrase the former
\jtcite{Jayn,Jayn2}.   What we  actually need to    do is to  maximise
$S_{G}$  subject   to  the constraints  imposed   by  our knowledge of
expectation  values of certain  operators $P_{1}[\phi, \partial \phi],
\ldots, P_{n}[\phi, \partial  \phi]$. In contrast to  equilibrium, all
$P_{k}[\ldots]$'s need not to be the constants of the motion. So namely if
one knows that 
 
\begin{equation}
\langle P_{k}[ \phi, \partial \phi] \rangle = f_{k}(x_{1}, x_{2}, \ldots),
 \jtleq{ivd}
\end{equation}

\noindent the entropy maximalisation leads to

\begin{displaymath}
\rho = 
\frac{e^{\left(-\sum_{i=1}^{n} \int \prod_{j}d^{4}x_{j} \;
\lambda_{i}(x_{1},\ldots)P_{i}[ \phi,
\partial \phi]\right)}}{Z[\lambda_{1}, \ldots, \lambda_{n}]}
\end{displaymath}

\noindent with the `partition function'

\begin{displaymath}
Z[\lambda_{1}, \ldots , \lambda_{n}] = 
\mbox{Tr}\left(e^{(-\sum_{i=1}^{n} 
\int \prod_{j}d^{4}x_{j} \;    
\lambda_{i}(x_{1},\ldots)P_{i}[ \phi,
\partial \phi])} \right).
\end{displaymath}

\noindent  It  is possible  to  show that  the  stationary solution of
$S_{G}$ is unique and maximal\jtcite{BM}.  In the previous cases the
time integration is not either present at all ($f_{k}$ is specified
only in the initial   time $t_{i}$), or  is  taken over the  gathering
interval $(-\tau , t_{i})$.

The Lagrange multipliers $\lambda_{k}$ might be eliminated if
one solves $n$ simultaneous equations

\begin{equation}
f_{k}(x_{1}, \ldots) = - \frac{\delta \; \mbox{ln}Z}{\delta
\lambda_{k}(x_{1},
\ldots )}.
 \jtleq{jg3}
\end{equation}

\noindent The explicit solution of (\ref{jg3}) may be formally written
as

\begin{equation}
\lambda_{k}(x_{1},\ldots ) = \frac{\delta \; S_{G}[f_{1},
\ldots , f_{n}]|{{}_{max}}}{\delta f_{k}(x_{1}, \ldots)}.
\jtlea{ent1}
\end{equation}

\section{Off-equilibrium dynamical equations}

In  this  section   we   derive  the   off-equilibrium Dyson-Schwinger
equations using the canonical formalism. We believe that this is a new
and     far   more  natural    formulation   for    systems  away from
equilibrium. The  more intuitive path-integral formulation of Calzetta
and Hu\jtcite{CH} is  not particularly suitable  in this case, because
the     connection    between kernels\jtcite{CH}  and    initial-time
constraints turns out to be rather non-trivial\jtcite{PJET}.

Let us deal first with a single  field $\phi$.  We start with
the action $S$ where $\phi$ is  linearly coupled to an external source
$J(x)$. For the  fields in the  Heisenberg picture, the operator
equation of motion can be written as 

\begin{equation}
\frac{\delta S}{\delta \phi(x)}[\phi = \phi^{J}] + J(x) = 0,
 \jtleq{4.0}
\end{equation}

\noindent where the index $J$ emphasises that $\phi$ is implicitly
$J$-dependent. It will prove useful in the following to reexpress
$\phi^{J}$ in such a way that the $J$-dependence will become explicit. The
latter can be done via an unitary transformation connecting  $\phi^{J}$
(governed by $H - J\phi$) with $\phi$ (governed by $H$). If $J(x)$ is
switched on at time $t=t_{i}$ we have

\begin{equation} T^{*}_{C}\left( \left\{ \frac{\delta S[\phi]}{\delta
\phi} +
J
\right\}\; \mbox{exp}(i \int_{C} d^{4}y \; J(y)\phi(y)) \right) = 0,
 \jtleq{4.12}
\end{equation}

\noindent with $T^{*}$ being the $T^{*}$-ordering. The close-time path
$C$ is the standard Keldysh-Schwinger path. Associating with the upper
branch of $C$ index `$+$' and with the lower one the index `$-$' one may
introducing  the metric $(\sigma_{3})_{\alpha \beta}$ ($\sigma_{3}$ is
the  usual Pauli matrix    and $\alpha, \beta   = \{+;-\}$)  and write
$J_{+}\phi_{+} -   J_{-}\phi_{-}    = J_{\alpha}\;(\sigma_{3})^{\alpha
\beta}\; \phi_{\beta} =     J^{\alpha}\;(\sigma_{3})_{\alpha  \beta}\;
\phi^{\beta}$. For the raised   and  lowered indices we   simply read:
$\phi_{+} =    \phi^{+}$ and $\phi_{-}   =  -\phi^{-}$  (similarly for
$J_{\alpha}$).   Taking   $\mbox{Tr}(\rho   \ldots  )$ with   $\rho  =
\rho[\phi, \partial \phi, \ldots]$, we get 

\begin{equation}
\frac{1}{Z[J]}\frac{\delta S}{\delta \phi(x)}\left[\phi^{\alpha}(x) =
- i\frac{\delta}{\delta J_{\alpha}(x)}\right]Z[J] = -J^{\alpha}(x) ,
\jtleq{4.15}
\end{equation}

\noindent with $Z[J]$ being   the generating   functional for
Green's functions.  Because     of  the $T^{*}$-ordering     the  time
derivatives could be pulled out of $[\ldots ]$. Eq.(\ref{4.15}) may 
equivalently be written as 

\begin{equation}
-J^{\alpha}(x) = \frac{\delta S}{\delta \phi(x)}\left[
\phi_{c}^{\alpha}(x)
 - i \frac{\delta}{\delta J_{\alpha}(x)} \right]\1.
\jtlea{4.16}
\end{equation}

\noindent The $\1$ indicates the unit. Analogously as for equilibrium
systems, we have defined the mean field $\phi_{c \; \alpha}(x)$ as

\begin{displaymath}
\phi_{c \; \alpha}(x) = \langle \phi_{\alpha}(x) \rangle_{J} =
(\sigma_{3})_{\alpha \beta}\frac{\delta
W[J]}{\delta
J_{\beta}(x)}, \;Z[J] = \mbox{exp}(iW[J]).
\end{displaymath}

\noindent Summation over contracted indices is understood. The effective
action $\Gamma [\phi_{c}]$ is connected with $W[J]$ via the Legendre
transform: $\Gamma[\phi_{c}] = W[J] - \int_{C} d^{4}y \; J(y) \phi_{c}(y)$.

With this mathematical setting we obtain the usual
equilibrium-like identities\jtcite{LW,KCC}

\begin{displaymath}
\frac{\delta \Gamma[\phi_{c}]}{\delta \phi_{c \; \alpha}(x)}
 = -J^{\alpha}(x),
\end{displaymath}
\begin{displaymath}
\int d^{4}y \;
G_{\alpha \beta}(x,y) \;(\sigma_{3})^{\beta \delta} \;\Gamma^{(2)}_{\delta
\gamma}(y,z) = (\sigma_{3})_{\alpha \gamma}\delta^{4}(x-z),
\end{displaymath}
\noindent with
\begin{displaymath}
- G_{\alpha
\beta}(x,y)  
= i \langle T_{C}\{ \phi_{\alpha}(x) \phi_{\beta}(y)\}
\rangle -i\langle \phi_{\alpha}(x)\rangle \langle \phi_{\beta}(y)\rangle
\;,
\end{displaymath}
\begin{displaymath}
\frac{\delta^{2}\Gamma}{\delta \phi^{\alpha}_{c}(x)
\delta \phi^{\beta}_{c}(y)}
= \Gamma^{(2)}_{\alpha \beta}(x,y) .
\end{displaymath}

\noindent For the physical process (i.e. $J_{\pm}=0$) we have from
(\ref{4.16})

\begin{eqnarray}
\frac{\delta \Gamma[\phi_{c}]}{\delta \phi_{c}^{\alpha}(x) } &=&
\frac{\delta 
S}{\delta \phi(x)}\left[
\phi_{c \; \alpha}(x) \right.\nonumber \\ 
&+& \left. i \int d^{4}y \; G_{\alpha
\beta}(x,y)\;(\sigma_{3})^{\beta \gamma}\;\frac{\delta}{\delta
\phi_{c}^{\gamma}(y)}\right] \1 = 0.  
\jtlea{4.20}
\end{eqnarray}  

\noindent It is worthy of noticing that the LHS of (\ref{4.20}) offers
a direct prescription for a calculation of $\delta \Gamma[\phi_{c}]
/\delta \phi_{c}^{\alpha}(x)$.

So far we have not  taken into  account  the constraints. This can  be
done quite simply. One  just sets $\lambda_{k}$ in  $\rho $ to be  the
solution of Eq.(\ref{jg3}). Using the identity 

\begin{displaymath}
e^{A}Be^{-A} = \sum_{n=0}^{\infty}\frac{1}{n!}C_{n},\;\;\;\; C_{0}=B,C_{n} = [A,C_{n-1}],
\end{displaymath}

\noindent we get then the generalised KMS condition

\begin{displaymath}
G_{+-}(x,y) = G_{-+}(x,y) +
\sum_{n=1}^{\infty}\frac{1}{n!}\mbox{Tr}(\rho \;\phi(x)C_{n}(y)).
\end{displaymath}

\noindent   Here  $A=    \mbox{ln}\rho$,     $B  =     \phi(y)$    and
$y_{0}=t_{i}$. Similarly   we  could    derive the   generalised   KMS
conditions for the higher point Green's functions 

Eq.(\ref{4.20}) and its  successive  $J$  variation provide us    with
the coupled integro-differential equations.  As a
result,  we get an infinite  hierarchy  of coupled equations which, if
furnished with   the corresponding  KMS   conditions, constitute,   in
principle,   a  complete   description     of the  behaviour    of   a
non-equilibrium system.

\section{The $O(N)\; \phi^{4}$ theory}

Let   us illustrate the    aforementioned  formalism on  the  $O(N)\;
\phi^{4}$ theory.

The $O(N)\; \phi^{4}$ theory is described by the bare Lagrange
function

\begin{equation}
{\cali{L}}= \frac{1}{2}\sum_{a=1}^{N}\left( (\partial
\phi^{a})^{2}-m_{0}^{2}(\phi^{a})^{2} \right) -
\frac{\lambda_{0}}{8N}\left( \sum_{a=1}^{N} (\phi^{a})^{2} \right)^{2}.  
 \jtleq{6}
\end{equation}

\noindent Using the explicit form (\ref{6}), Eq. (\ref{4.20})
reads

\begin{eqnarray*}
&&\frac{\delta \Gamma}{\delta \phi^{a\; \alpha}(x)} = -(\Box +
m_{0}^{2})\phi_{\alpha}^{a}(x)\\
&&\\
&&\;\; -\frac{\lambda_{0}}{2N}\left\{
\sum_{b=1}^{N}\phi_{\alpha}^{a}(x)(\phi_{\alpha}^{b}(x))^{2}+
i\phi_{\alpha}^{a}(x)\sum_{b=1}^{N} G^{bb}_{\alpha
\alpha}(x,x)\right.\\
&&\;\;+ i2
\sum_{b=1}^{N}\phi^{b}_{\alpha}(x)G^{ab}_{\alpha \alpha}(x,x)\\
&&\\
&&\;\;+
\left. \int d^{4}y\; d^{4}w \;d^{4}z \; \sum_{b=1}^{N} G^{(3)\;abb}_{\alpha
\alpha \alpha}(y,w,z)\right\} = 0. 
\end{eqnarray*}

\noindent A successive variation with  respect to $J(y)$ generates
the DSE for the two-point Green's functions. 

The dynamical equations can be significantly simplified provided that
both the density matrix and the Hamiltonian are invariant against rotation
in the $N$-dimensional vector space of fields. This fact leads
straightforwardly to the following Ward's identities\jtcite{PJET}

\begin{eqnarray*}
\left.\frac{\delta W[J]}{\delta J^{a}_{\beta}(z)}\right|_{J=0} &=&
\phi^{\beta \;
a}(z)
= 0,
\;\;\;\;\;\; \forall a,\\
&&\\
\left. \frac{\delta^{2}W[J]}{\delta J^{\alpha\;a}(x) \delta
J^{\beta\;b}(y)}\right|_{J=0} &=& G^{ab}_{\alpha \beta}(x,y) = \delta^{ab}G_{\alpha \beta}(x,y),\\
&&\\
\Gamma^{(3) \;abc}_{\alpha \beta
\gamma}(y_{1}, y_{2}, y_{3})&=& 0, \;\;\;\; \forall a,b,c,\\
&&\\
\Gamma^{(4)\; abcd}_{\alpha \beta
\gamma \delta}(y_{1}, y_{2}, y_{3}, y_{4})&\propto & \; \delta_{ab}\delta_{cd} +
\delta_{ac}\delta_{bd} +
\delta_{ad}\delta_{bc}. 
\end{eqnarray*}

\noindent With  these results we  may write the evolution equation for
the two-point Green's  function  as follows   

\begin{eqnarray}
&&\left(\Box + m_{0}^{2} +
\frac{\lambda_{0}}{2}\;i \;\frac{N+2}{N}\;G_{\alpha \alpha}(x,x)
\right)G_{\alpha
\beta}(x,y)  \nonumber \\
&&\nonumber \\
&&+ \;\frac{\lambda_{0}}{2}\;\frac{N+2}{N}
\left( G^{(4)}_{\alpha \alpha \alpha \beta}(x,x,x,y)\right)= \; - \delta(y-x)(\sigma_{3})_{\alpha \beta}, 
\jtlea{me1}
\end{eqnarray}

\noindent In the following we shall confine ourselves only to such
situations where both $\rho$ and $H$ are $O(N)$ invariant.

\section{$G_{\alpha \beta}(x,y)$: $O(N)\; \phi^{4}$ theory in the
large $N$ limit}

One can show \jtcite{M1,M2,PJET} by a detailed study of the large-N
approximation (virtually using only the Ward's identities and
properties of $\Gamma$ an $W$) that the vertex functions $\Gamma^{(2n)}$ must be of order $N^{1-n}$.  The
latter suggests that in the dynamical equation (\ref{me1}) we can
neglect $\Gamma^{(4)}$ terms.

Let us mention one more point. If we  perform the expectation value of
the Lagrange function (\ref{6})  we find that this  does not depend on
$G^{(4)}$ in the $N \rightarrow \infty$ limit, indeed 

\begin{eqnarray*}
\mbox{Tr}(\rho \; {\cali{L}}(x)) 
&=& \frac{1}{2} iN \;\{\partial_{\alpha\; x} \partial^{\alpha}_{y} G(x,y)
|_{x=y} - m_{0}^{2} G(x,x)\} \nonumber \\
&& \nonumber \\
&-& \frac{i\lambda_{0}\;(N + 1)}{4} \{G^{(4)}(x,x,x,x) + (G(x,x))^{2}\}.
\end{eqnarray*}

\noindent  Here  we have    used  the  fact  that  $G^{(4)}_{aaaa}   =
3G^{(4)}_{aabb} = 3G^{(4)}$. So we may  equally start with the Lagrange
function 

\begin{eqnarray}
{\hat{\cali{L}}} &=& \frac{1}{2} \sum_{a=1}^{N} \left( (\partial  
\phi^{a}(x))^{2} - m_{0}^{2}(\phi^{a}(x))^{2} \right)\nonumber \\
&-&
\frac{\lambda_{0}}{4} \sum_{a=1}^{N} (\phi^{a}(x))^{2}G(x,x).
\end{eqnarray}

\noindent The former fulfils the identity $\langle {\hat{\cali{L}}} \rangle =
\langle {\cali{L}} \rangle |_{N \rightarrow \infty}$. It is also simple to
see that the DSE derived from
${\hat{\cali{L}}}$ reads

\begin{eqnarray}
&&\left( \Box_{x} + m_{0}^{2} + \frac{i \lambda_{0}}{2} \; G_{\alpha
\alpha}(x,
x) \right) G_{\alpha \beta}(y,x)\nonumber \\
&& = -\delta(x-y) (\sigma_{3})_{\alpha
\beta}.
\jtlea{DSE1}
\end{eqnarray}

\noindent This is precisely the same which one would obtain
if the large $N$ limit would be performed in the original DSE (\ref{me1}).

Let us now show how to compute $G_{\alpha \beta}(x,y)$ for some familiar
choices of the initial-time constraints.

\vspace{3mm}  

\noindent{\em (i) Equilibrium}

\vspace{2mm}

In this case the constraints are usually chosen to be the integrals
of motion. The only available integral of the  motion is the full
Hamiltonian $H$, and the corresponding constraint reads  \jtcite{Tol,Cub}

\begin{equation}
\langle P_{k}[\phi, \partial \phi] \rangle|_{t_{i}} = \langle H \rangle =
\int_{0}^{T}dT' C_{V}(T') = F(T),
\end{equation}

\noindent  where  $C_{V}$  is the heat   capacity  at constant  volume
$V$.   The density matrix   $\rho$ maximising $S_{G}$  is the 
density    matrix    of   the      canonical   ensemble:    $\rho    =
\frac{\mbox{exp}(-\beta H)}{Z[\beta]}$. The Lagrange multiplier $\beta$
is determined from Eq.(\ref{ent1}):

\begin{equation}
\beta = \frac{\partial S_{G}}{\partial F(T)} = \left( \frac{\partial
S}{\partial T }\right)_{V}\left(\frac{\partial F(T)}{\partial
T}\right)^{-1}_{V} = \frac{1}{T}.
\end{equation}

\noindent In this case the KMS condition is the well known relation

\begin{equation}
G_{+-}({\vect{x}},t_{i}; {\vect{y}},0) =
G_{-+}({\vect{x}},t_{i}-i\beta; {\vect{y}},0).
\end{equation}

\noindent The DSE are those in (\ref{DSE1}) with $G(x,y)=G(x-y)$.
The solutions are the equilibrium propagators
in the Keldysh-Schwinger formalism, i.e.

\begin{eqnarray}
iG_{\pm \pm}(k) &=& \frac{\pm i}{k^{2} - m^{2}_{r} \pm i\varepsilon} +
2\pi f(|k_{0}|)\delta(k^{2}-m^{2}_{r})\nonumber \\
iG_{\pm \mp}(p) &=& 2\pi \left\{ \theta(\mp k_{0}) + f(|k_{0}|)
\right\}\delta(k^{2} 
- m^{2}_{r})\; ,
\jtlea{sol1}
\end{eqnarray}

\noindent where $f(x) = (\mbox{exp}(\beta x)-1)^{-1}$ is the Bose
distribution, and  $m^{2}_{r
} = m^{2}_{0} + \frac{i\lambda_{0}}{2}G(0)$ is the renormalised mass. 

\vspace{4mm}

\noindent {\em (ii) Non-equilibrium: translationally invariant $G_{\alpha
\beta}$}

\vspace{2mm}

The  DSE in the former  example were immensely  simplified  due to the
translational invariance of   Green's functions.   If  we retain   the
translational invariance this simplicity  will  be preserved also   to
non-equilibrium.   As an  example,   let    us choose the    following
initial-time constraint: 

\begin{equation}
\langle P[\phi , \partial \phi] \rangle |_{t_{i}} = \langle {\cali{H}}
({\vect{k}}) \rangle = \langle
{\hat{\cali{H}}}({\vect{k}}) \rangle = g({\vect{k}}),
 \jtleq{DS5} 
\end{equation}

\noindent where ${\hat{\cali{H}}}$ means the effective Hamiltonian
density derived from ${\hat{\cali{L}}}$. The density matrix reads

\begin{displaymath}
\rho = \frac{\mbox{exp}(-\int d^{3}{\vect{k}} \beta({\vect{k}})
{\hat{\cali{H}}}({\vect{k}}) )}{Z[\beta]} = \frac{\mbox{exp}(- \int
d^{3}{\vect{k}}
{\tilde{\beta}}({\vect{k}})a^{\dagger}_{\vect{k}}
a_{\vect{k}})}{Z[\beta]},
\end{displaymath}

\noindent        with           $\beta({\vect{k}})                   =
{\tilde{\beta}}({\vect{k}})2\sqrt{{\vect{k}}^{2}   +  m^{2}_{r}}$  and
$\beta({\vect{k}})  = \frac{\delta S_{G}}{\delta g({\vect{k}})}$.  The
former indicates   that    the  different  modes   have   different
`temperatures'.  The DSE agree with those in the equilibrium case.  The
generalised KMS conditions are 

\begin{equation}
G_{+-}(k) = e^{-{\tilde{\beta}}({\vect{k}})k_{0}/2}G_{-+}(k) 
\jtleq{KMS1}
\end{equation}

\noindent The solution of (\ref{DSE1}) furnished with (\ref{KMS1})
formally coincides with the solution (\ref{sol1}). The only proviso is
that $f(|k_{0}|) \rightarrow \frac{1}{e^{\beta({\vect{k}})/2}-1}$.

\section{Conclusions and outlook}

We considered the  Jaynes-Gibbs  principle of maximal  entropy.   This
allowed    us  to  construct    the   non-equilibrium  DSE.   For  the
$O(N)\;\phi^{4}$  theory in the  $N  \rightarrow \infty$ limit we have
explicitly evaluated propagators  for two  choices of  translationally
invariant  density matrices. A  notable advantage  of this approach is
that  one   can   get     the   DSE without    going      through  the
Cornwall-Jackiw-Tomboulis formalism\jtcite{CHKMP,M1,M2} which would be
formidably difficult particularly for more than one constraint. 

The $O(N)\; \phi^{4}$ theory in   the large $N$ limit   is a nice  toy
model  allowing in many   cases to  approach  the dynamical  equations
analytically.  For  suitable   choices    of   the translationally
non-invariant initial-time  constraints the simplicity of the equations
is such that one may solve them exactly. A more detailed report will
appear elsewhere\jtcite{PJET}.

\section{Acknowledgements}

This work is supported by Fitzwilliam College and CONACYT.


\begin{references}


\bibitem{Jayn} E.T.Jaynes, Am.J.Phys. 33 (1965) 391.

\bibitem{Jayn2} E.T.Jaynes, Phys.Rev 106 (1957) 620, 108 (1957) 171.

\bibitem{Tol} R.C.Tolman, The Principles of Statistical Mechanics
(Clarendon Press, Oxford, 1938).

\bibitem{makhlin} A.Makhlin, Phys.Rev.C 51, (1995) 3454.

\bibitem{CHKMP} F.Cooper, S.Habib, Y.Kluger, E.Mottola and J.P.Paz,
hep-ph/9405352.

\bibitem{M1} F.Cooper and E.Mottola, Phys.Rev.D 36 (1987) 3114.   

\bibitem{M2} F.Cooper, S.Habib, Y.Kluger and E.Mottola, Phys.Rev.D 55
(1997) 6471.

\bibitem{EJY} O.{\'E}boli, R.Jackiw and So-Young Pi, Phys.Rew.D 37
(1988) 3557.

\bibitem{Sh} C.E.Shannon and W.Weaver, The Mathematical Theory of
Communication (University of Illinois Press, Urbana, 1949).

\bibitem{Gibbs} J.W.Gibbs, Elementary Principles in Statistical
Mechanics (Dover Publications, New York, 1960).

\bibitem{BM} B.Buck and V.A.Macaulay, Maximum Entropy in Action
(Oxford, 1991). 

\bibitem{CH} E.Calzetta and B.L.Hu, Phys.Rev.D 37 (1988) 2878. 
 
\bibitem{PJET} P.Jizba and E.S.Tututi, {\em The Jaynes-Gibbs principle
of maximal entropy and the non-equilibrium dynamics of the $O(N)\;
\phi^{4}$ theory at large $N$}, to appear.  

\bibitem{LW} N.P.Landsman and Ch.G.van Weert, Phys. Rep. 145,
(1987)  141.

\bibitem{KCC} K.C.Chou, Z.B.Su, B.L.Hao and L.Yu, Phys. Rep. 118
(1985) 1.

\bibitem{Cub} R.Kubo, Statistical Mechanics (North-Holland Publishing
Company, Oxford, 1965.






\end{references}
\end{document}